\documentclass[sigconf]{acmart}

\author{Dino Conciatore}
\affiliation{
  \institution{Swiss National Supercomputing Centre, ETH Zurich}
  \city{Lugano}
  \country{Switzerland}
}

\author{Elia Oggian}
\affiliation{
  \institution{Swiss National Supercomputing Centre, ETH Zurich}
  \city{Lugano}
  \country{Switzerland}
}

\author{Federico Da Forno}
\affiliation{
  \institution{Swiss National Supercomputing Centre, ETH Zurich}
  \city{Lugano}
  \country{Switzerland}
}

\author{Stefano Schuppli}
\affiliation{
  \institution{Swiss National Supercomputing Centre, ETH Zurich}
  \city{Lugano}
  \country{Switzerland}
}

\author{Jerome Tissieres}
\affiliation{
  \institution{Swiss National Supercomputing Centre, ETH Zurich}
  \city{Lugano}
  \country{Switzerland}
}

\author{Joost VandeVondele}
\affiliation{
  \institution{Swiss National Supercomputing Centre, ETH Zurich}
  \city{Lugano}
  \country{Switzerland}
}

\author{Maxime Martinasso}
\affiliation{
  \institution{Swiss National Supercomputing Centre, ETH Zurich}
  \city{Lugano}
  \country{Switzerland}
}

\usepackage{enumitem}
\usepackage{hyperref}

\usepackage{graphicx} 
\usepackage{listings} 
\usepackage{xcolor}
\usepackage{url}
\usepackage{cleveref}

\title{Beyond Pre-Training: The Full Lifecycle of Foundation Models on HPC Systems}

\acmConference[CUG '26]{Cray User Group}{April 26-30, 2026}{Nice, France}

\begin{document}

\begin{abstract}
Large-scale pre-training of Foundational Models (FM) constitutes a computationally intensive first phase for enabling AI across diverse scientific and societal applications. This first phase has positioned High-Performance Computing (HPC) facilities as indispensable backbones of ``Sovereign AI'' initiatives. While the massive throughput requirements of FM pre-training align with the traditional capability-oriented mission of HPC, subsequent phases of the AI lifecycle, typically referred to as fine-tuning and inference, introduce operational paradigms that can conflict with established batch-processing environments. Moreover, these phases are not computationally trivial: they often require substantial high-end compute resources while exhibiting hardware utilization patterns that differ significantly from those of pre-training.

This paper addresses the architectural and strategic challenges of operationalizing a complete AI lifecycle within a national supercomputing facility. We present a hybrid cloud-native platform being developed and deployed at the Swiss National Supercomputing Centre (CSCS) that combines diskless GPU-enabled HPE Cray EX compute nodes with virtualized commodity infrastructure. Orchestrated by Kubernetes, this novel service architecture bridges the gap between HPC batch processing and service-oriented workflows. We report our initial investigations into fine-tuning pipelines and highly available inference services, analyzing the associated trade-offs while improving user productivity. Our findings offer a blueprint for enabling supercomputers to integrate ``AI Factories'' services and workflows, supporting AI innovations into end-to-end scientific and industrial use cases.
\end{abstract}

\maketitle

\section{Introduction}

The proliferation of Large Language Models (LLMs) represents the most visible facet of a broader adoption of Machine Learning (ML) Foundation Models (FMs) across society and science. Beyond natural language, data-driven approaches are increasingly augmenting traditional methodologies across a wide range of domains. For example, in weather and climate science \cite{WeatherGenerator2025}, they provide a complementary paradigm to conventional numerical simulations. This technological adoption has prompted governments to pursue sovereign AI capabilities, aiming not only at regulatory compliance and cultural representation, but also at strategic independence of nation-critical computational and informational assets.

Large models development such as the Apertus model~\cite{apertus} inherently requires state-of-the-art HPC resources because of the computational intensity and data scale involved, especially in their early stages. Consequently, for sovereign and non-commercial initiatives, it is natural to approach national facilities such as the \textit{Alps} Research Infrastructure (RI) operated by the Swiss National Supercomputing Centre (CSCS) in Lugano, Switzerland. These represent the natural ecosystem for such undertakings, providing both the capital-intensive hardware, primarily represented by massive GPU partitions, high-end interconnects, and suitable storage systems, as well as the world-class engineering talent required to execute such specialized programs.

Although Foundation Models pre-training aligns with the \\ \textit{capabilities-oriented} mission of HPC facilities (i.e., specialized infrastructures dedicated to enable otherwise intractable computational problems), pre-training alone is insufficient for developing AI sovereignty. Realizing the value of these models requires, among other things, post-training adaptation (fine-tuning) and deployment (inference). However, these downstream tasks introduce a conflict: while fine-tuning may still demand significant infrastructure resources, inference often entails high-capacity workloads of independent instances that do not strictly require a supercomputer architecture. This creates tension between the typical \textit{capabilities-oriented} mandate and the heterogeneous requirements of the typical AI project lifecycle, raising a strategic dilemma: should HPC facilities broaden their operations to support these workflows, or is this a substantial deviation from their core mission?

The necessity to resolve this dilemma is heightened by the emerging European framework of \textit{AI Factories} (AIFs) and their associated \textit{Antennas}~\cite{EU_AIfactories}. These initiatives aim to democratize access to supercomputing, extending the infrastructure's reach beyond traditional academic research to include key industrial sectors where AI can have transformative impact, such as Life Sciences \& Healthcare, Weather Prediction, as well as enabling SMEs and startups. To fulfill this vision, HPC facilities have to support the research and development activities of these diverse stakeholders in a secure and cost-effective way with the least amount of friction possible.

Accommodating this broader community introduces distinct architectural challenges. Regarding fine-tuning, while current workflows are Slurm-centric, we anticipate a shift toward demand for ``one-click'' solutions, particularly from new user communities (including, SMEs involved in R\&D) who may lack either the capabilities or the time to deal with the intrinsic complexity of HPC systems, and will therefore require higher levels of abstraction. However, implementing such level of abstraction is non-trivial; fine-tuning involves multiple stages (e.g., SFT, RL) and techniques (e.g., LoRA) that require proper calibration to prevent damaging the underlying model capabilities \cite{FT_can_damange_the_model}. Regarding inference, the need for long-running, highly available services conflicts with the standard batch-processing model. Emulating such features via job schedulers is operationally brittle, relying on, e.g., resubmission workarounds. Furthermore, long-running inference workloads create a distinct optimization challenge: preventing high-end HPC nodes from idling during low-traffic periods. Crucially, these integration hurdles are not exclusive to LLMs, but apply equally to other scientific domains where the software product ecosystem is often not yet as established.

In response, CSCS is actively exploring architectural solutions to address these challenges. This effort is anchored in a co-design process involving several key representative groups, who have been granted early access to help define the necessary service models.
Therefore, our work presented in this paper provides a technical and operational framework for bridging the gap between high-performance ``capability'' computing and cloud-native ``service'' environments in the context of AI services. In particular, our contributions are:

\begin{itemize}
    \item A hybrid architecture for an AI Factory that integrates diskless HPE Cray EX nodes and commodity hardware (VMs) into one Kubernetes-managed control plane.
    \item A methodology for operationalizing the ``Full AI Lifecycle'', extending HPC services from the batch model used for pre-training, to a set of services suitable for supporting fine-tuning and high-availability inference.
    \item A set of initial performance results together with technical trade-off of running distributed training and high-throughput inference within a containerized, service-oriented ecosystem on Slingshot-interconnected hardware.
\end{itemize}

This paper is thus organized as follows. After further expanding on the core motivations in \hyperref[sec:motivations]{Section~\ref*{sec:motivations}}, and reviewing related work in \hyperref[sec:related_approaches]{Section~\ref*{sec:related_approaches}}, we outline the ongoing efforts to provide end-to-end AI lifecycle support within our HPC context in \hyperref[sec:tech-approach]{Section~\ref*{sec:tech-approach}}. By presenting our initial findings and the architectural trade-off encountered, our aim is to contribute to the solution on how HPC centres can empower a sovereign AI ecosystem, and ultimately define a blueprint of those solutions to share across the HPC community. We conclude by discussing results obtained so far in \hyperref[sec:results]{Section~\ref*{sec:results}}, ongoing efforts and anticipated future work in \hyperref[sec:ongoing_future_work]{Section~\ref*{sec:ongoing_future_work}}.

\section{Motivations}
\label{sec:motivations}

The strategic dilemma outlined in the introduction, whether HPC facilities should expand beyond their traditional \textit{capability-oriented} mandate, relies on a historical dichotomy between ``batch processing'' and ``interactive services.'' However, as AI advances to become intrinsic to the scientific method, this dichotomy is increasingly difficult to sustain. The motivation to support the full AI lifecycle at CSCS is driven not merely by customer demand, but by the necessity to support modern scientific workflows and the operational efficiency of the infrastructure itself.

A primary example of this shift is found in numerical weather prediction (NWP). For decades, CSCS has provided computational capability for weather forecasting through precise, deterministic production workflows with rigid deadlines to meet product generation operational requirements. As AI-driven weather models evolve, the paradigm is shifting from these fixed-interval simulations toward the deployment of foundational models specialized for regional domains. This evolution fundamentally alters the computational cost and architectural requirements: in this new regime, weather prediction may transition to on-demand inference services tailored to specific locations and times. Such services can generate tens of terabytes of data within minutes; at scale, a high volume of concurrent requests risks a ``data explosion'' and severe I/O bottlenecks that only HPC-class parallel file-systems and high-bandwidth interconnects can mitigate.

\subsection{Enabling the Full AI Lifecycle}
\begin{figure}[h]
    \centering
    \includegraphics[width=\columnwidth]{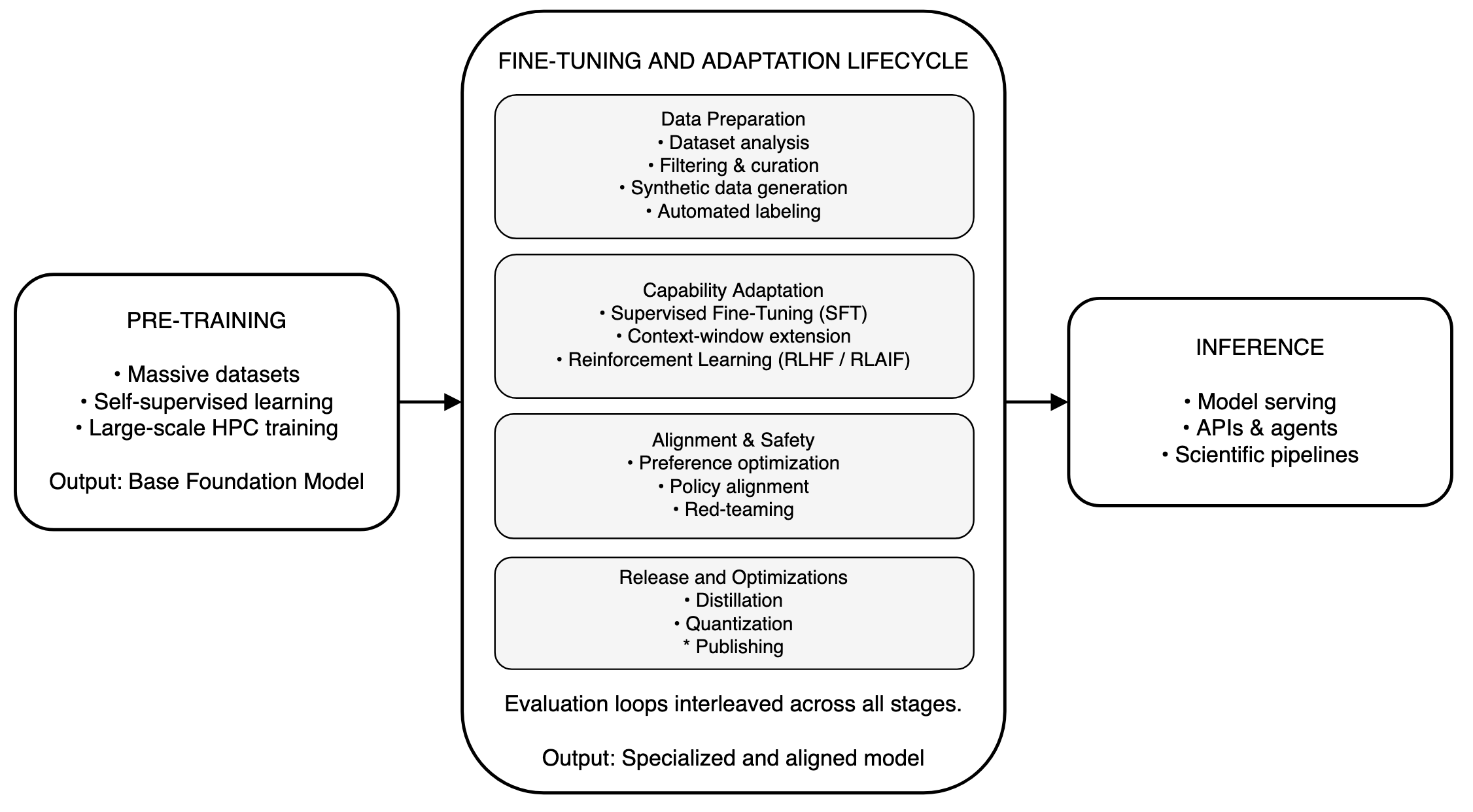}
    \caption{Lifecycle of foundation models in scientific environments. Large-scale pre-training produces a base model whose capabilities are subsequently specialized through an iterative post-training adaptation process---including datasets preparation, fine-tuning, alignment, safety validation, and release optimizations, before publishing and/or deployment for inference.}
    \label{fig:workflow}
\end{figure}

The pursuit of \textit{Sovereign AI} may be oversimplified as merely the construction of model weights. Effective autonomy requires control over the entire development and deployment pipeline. Without on-premise service infrastructure, academic and industrial researchers may be required to export pre-trained models to commercial clouds for post-training adaptation, which can introduce technical friction and decouple the model from its primary data source. By operationalizing the lifecycle depicted in \Cref{fig:workflow}, CSCS aims to support and enable the fulfillment of these diverse needs entirely within the national infrastructure. This approach allows users to manage the entire process, from data preparation to deployment, while leveraging the performance of the Cray EX fabric.

The model lifecycle illustrated in \Cref{fig:workflow} highlights that while pre-training establishes broad capabilities, scientific and industrial utility is primarily realized during the post-training adaptation phase. However, the sequence of activities required to refine a foundation model, spanning data preparation, supervised fine-tuning, alignment, and safety validation, presents a significant technical barrier for many potential users. For SMEs and specialized research team, the expertise required to navigate these adaptation steps can hinder the exploration of new research directions. Consequently, there is a need to facilitate this process for non-expert users by providing an integrated infrastructure that simplifies the underlying technical complexity. This need is further motivated by data gravity, where moving petabyte-scale datasets to external clouds is often untenable, and by the requirement for low-latency interconnects for bespoke distributed training methods. By enabling integrated orchestration on-premise, HPC facilities can bridge these gaps and support the full adaptation chain without the latency or security concerns associated with external network boundaries. This comprehensive workflow is discussed in further detail in \Cref{sec:tech-approach}.

\subsection{Agility through Autonomy}
\label{sec:agility_through_autonomy}
The ML community operates in a landscape where innovation and trends can shift overnight. To maintain a competitive edge, researchers and SMEs require the agility to move fast, which in turn necessitates a high degree of operational autonomy. Traditional HPC support models, where users may wait for staff to deploy specific services, cannot scale to meet the diverse and rapidly evolving requirements of this community. Furthermore, a centralized engineering team does not have sufficient capacity to handle the extensive and continually changing demands for software products and specialized workflows needed by various user groups.

To address this, CSCS is developing the concept of ``self-managed sandboxes''. We aim to provide a set of general, robust tools and a catalog of documented blueprints that allow users to build and manage their own custom workflows with the software stack of their choosing. We believe this autonomy provides a significant advantage in a highly competitive field, enabling users to iterate on R\&D directions without being impeded by centrally-managed service architectures. For users primarily focused on model consumption rather than adaptation, we have also developed a managed model inference service that provides standardized, scalable inference endpoints for both public and custom models.

While the work presented in this paper concentrates on the technical integration of these workflows, we recognize that working with sensitive data demands additional privacy frameworks. Our efforts are thus designed to integrate with the privacy-preserving, air-gapped environments we are currently developing at CSCS.

\section{Overview of Related Approaches}
\label{sec:related_approaches}

The computational needs of FMs have traditionally been separated by phase. Large-scale pre-training, with massive data parallelism and synchronous communication, has fit naturally on HPC clusters managed by queue-based workload managers. But the rise of “AI Factories” and similar sovereign efforts has broadened the focus to the full AI lifecycle, extending beyond pre-training. As a result, the available literature increasingly examines service architecture questions to better meet the specific needs of fine-tuning, inference, and agentic workloads.

To bridge the gap to traditional HPC, meta-orchestrators like ColonyOS \cite{colonyos} and scientific workflow systems such as RADICAL-Pilot \cite{merzky2025radical} have been developed. These frameworks typically treat the supercomputer as a remote execution backend, abstracting the complexity of the batch scheduler while maintaining a logical separation between the service layer and the compute fabric. Complementary to these are efforts such as the \textit{Scalable Engine}  \cite{trappen2025automated} proposed for the RAMSES supercomputer, which integrates vLLM with Slurm and Kubernetes to provide an automated scaling mechanism for bursty inference demands within a batch environment.

Within the Swiss ecosystem, \textit{OpenTela} \cite{yao2025opentela}, part of the Swiss AI Initiative’s serving platform, enables scalable LLM serving by allowing researchers to multiplex requests to shared model instances deployed on platforms such as Slurm without requiring administrative privileges. The resulting serving platform was then extended with \textit{Model Spinning} \cite{model_spinning}, a solution based on CI/CD pipelines and the FirecREST API \cite{palme2025firecrestv2lessonslearned} to ensure that "hot" models are always available for inference while backed by Slurm jobs.

The necessity of these heterogeneous workflows is summarized by Garcia Lopez et al. \cite{garcia2025aifactory}, who advocate for a ``dual-stack'' architecture where HPC and cloud-native platforms share accelerators and data more intelligently. Their work highlights the need for infrastructure capable of simultaneously supporting batch-oriented HPC workloads and service-oriented AI systems, which can exhibit long-lived, interactive, or bursty resource usage patterns.

Another line of work investigates the direct convergence of Kubernetes and HPC schedulers. For example, High Performance Kubernetes (HPK) \cite{chazapis2023hpk} enables the execution of unmodified Kubernetes workloads on HPC clusters by translating Kubernetes pod lifecycle operations into Slurm jobs executed through Singularity/Apptainer containers. In this model, Kubernetes serves primarily as a user-facing orchestration layer, while the HPC scheduler remains the authoritative resource manager. Similar approaches embed HPC runtimes within Kubernetes environments or create virtual clusters where Slurm controls resource allocation while Kubernetes manages containerized services. While these approaches successfully expose cloud-native tools to HPC environments, they generally preserve the traditional HPC control-plane hierarchy.

In contrast, the approach presented in this work uses Kubernetes as an alternative compute plane that unifies commodity hardware (VMs) and HPC nodes into a single platform. This provides end users with an environment for workloads ranging from fine-tuning to inference, and is well suited for custom and advanced setups (e.g., agentic-centric services).

\section{Our Technological Approach}
\label{sec:tech-approach}

To enable the full AI lifecycle on the \textit{Alps} Research Infrastructure, we implement a hybrid architecture orchestrated via Kubernetes \cite{evolvingHPCforML}. This platform unifies high-end diskless HPE Cray EX compute nodes with virtualized commodity hardware into a single, cohesive compute plane. Our approach moves beyond the traditional ``remote backend'' model used in scientific workflow systems, where HPC clusters are treated as external execution engines. Instead, the supercomputing fabric becomes an integrated component of a cloud-native orchestration environment.

Unlike approaches where Kubernetes workloads are translated into HPC batch jobs, our design places Kubernetes as the primary orchestration layer and integrates HPC nodes directly into the cluster as schedulable resources. In this model, Slurm continues to operate independently for large-scale pre-training and tightly coupled simulations, while service-oriented workloads such as inference pipelines, agentic systems, and RAG applications are orchestrated natively through Kubernetes.

A distinguishing factor of our design is its domain-agnostic nature. While LLMs benefit from a rich and evolving ecosystem of software products (e.g., vLLM, LiteLLM), the primary scientific communities served by CSCS, including Numerical Weather Prediction and medical research, might also rely on models that are not LLM-centric. The provision of self-managed sandboxes serves as our primary tool for achieving this domain-agnostic nature, allowing users to deploy bespoke software stacks that are not purely focused on LLMs.

Furthermore, we recognize that users, particularly SMEs, might require the flexibility to eventually migrate their R\&D products from a national facility to commercial cloud providers. For service-oriented AI products, as opposed to traditional batch simulations, the destination is almost invariably a Kubernetes environment. By exposing Kubernetes namespaces directly to users on the CSCS infrastructure, we significantly reduce or eliminate potential \textit{vendor lock-in}, facilitating a seamless transition to external platforms when project requirements evolve.

The resulting infrastructure (referred to as \textit{Alpernetes}) combines \textit{Alps} nodes, leveraging vCluster technology \cite{vclusters_24}, with commodity hardware Virtual Machines. Currently, about a dozen \textit{Alps} nodes have been dedicated to this purpose, representing a small part of the 2,500+ node system. The core of our offering consists of three primary service tiers designed to provide the flexibility required by both LLM and non-LLM communities:

\begin{itemize}

\item \textbf{Self-Managed Sandboxes:} We provision namespace-isolated environments where users gain access to a \textbf{unified compute plane} spanning both commodity hardware (VMs) and high-end HPC nodes. This heterogeneity allows researchers to build complex service stacks, utilizing GPU resources for models hosting while leveraging commodity hardware for integration services, web interfaces and databases. These sandboxes are ideal for developing RAG and agentic services, providing an environment suitable for rapid iterative experimentation needed in the rapidly evolving ML field. Crucially, they empower non-LLM communities to autonomously deploy custom software stacks that do not fit into mainstream templates.

\item \textbf{Fine-tuning as a Service:} We are currently in an exploratory phase focused on defining the requirements for managed fine-tuning pipelines. We aim at abstraction levels that cater to both ends of the user base spectrum: expert users on one side who will likely prefer to continue running bespoke fine-tuning activities directly via Slurm for maximum control, and on the other end, user groups that require a simplified "one-click" experience. For the latter, the goal is to provide a high-level interface where they can simply select a base model from a curated catalog, upload their dataset, and initiate the fine-tuning process through an automated workflow we established.

\item \textbf{Managed Inference Service:} CSCS utilizes the sandbox logic to deploy persistent inference endpoints for widely used models. These require a mix of internet-facing APIs and high-throughput engines like vLLM running on \textit{Alps} GPUs, ensuring that models are accessible via standardized protocols for downstream scientific applications. This service cater to user groups who are only interested in LLM-based inference endpoints.

\end{itemize}

Conceptually, the resulting architecture forms a dual-control-plane system where batch-oriented HPC workloads continue to be orchestrated by Slurm, while part of the service-oriented AI workloads are managed through Kubernetes. Each of the above topics will be discussed in detail in the remainder of this chapter.

\subsection{Kubernetes Clusters}

Our private cloud infrastructure is designed to provide a unified operational model for conventional cloud services, data-intensive applications, and high-performance computing (HPC) and AI workloads. The platform combines virtualization, Kubernetes lifecycle management, GitOps, and infrastructure-as-code in order to provision clusters consistently across heterogeneous compute resources.

At the control-plane level, \textbf{SUSE Rancher} \cite{rancher} is used as the central platform for managing Kubernetes clusters. Rancher provides a single operational interface for cluster registration, lifecycle management, access control, and multi-cluster governance. To supply virtualized resources dynamically to Rancher, we rely on \textbf{SUSE Virtualization} (formerly Harvester) \cite{suse_virtualization}, which provides a Kubernetes-based hyperconverged infrastructure platform for running virtual machines and containers on bare-metal servers.

Deployment and configuration are handled according to a \textbf{GitOps} model. \textbf{ArgoCD} \cite{argocd} is used with the \emph{app-of-apps} pattern to manage all cluster deployments in a declarative and reproducible way. In practice, both platform services and cluster-specific components are defined in Git and continuously reconciled into the target environments. This approach ensures that clusters remain aligned with the desired state and that operational changes remain traceable and auditable.

\subsubsection{Infrastructure as Code and Environment Management}

The infrastructure is described using \textbf{OpenTofu} \cite{opentofu}. Each Kubernetes cluster is represented by a dedicated folder in a repository containing its manifests and variables. Through this structure, it is possible to define the number of master and worker nodes, CPU and memory sizing, disk size, VLAN assignment, container network interface (CNI), and cluster ownership through OIDC groups.

For VM-based clusters, nodes are assigned automatically as part of the provisioning process. For pure bare-metal clusters, OpenTofu generates the cluster definition, while node enrollment is completed later through Ansible-based workflows.

OpenTofu is also used to define which baseline platform components must be deployed on each cluster. These include common services such as \textbf{MetalLB}, \textbf{external-dns} with \textbf{Let's Encrypt}, \textbf{CSI drivers}, \textbf{cert-manager}, and \textbf{external-secrets} integrated with \textbf{Vault} \cite{vault}. Once declared in OpenTofu, these components are enabled through ArgoCD ApplicationSets so that they are deployed in the same standardized way as all other cluster services. This provides a consistent bootstrap process across environments and cluster types.

To simplify orchestration of the different OpenTofu modules, we use \textbf{Terragrunt} \cite{terragrunt} as a wrapper layer. Terragrunt bundles the infrastructure definitions, reduces code duplication, and provides a clear separation across three environments: \textbf{production}, \textbf{test/staging (TDS)}, and \textbf{development}. This organization makes it possible to promote infrastructure changes progressively while preserving a uniform operational model.

\subsubsection{Compute Node Flavours}

The infrastructure supports four main compute-node flavours, each addressing different operational and performance requirements.

\paragraph{SUSE Virtualization VMs}

The first flavour consists of virtual machines running on \textbf{SUSE Virtualization} over commodity hardware. These nodes provide the most flexible and operationally simple compute substrate. They are well suited for control-plane services, lightweight worker pools, development clusters, web applications, APIs, databases, internal services, and other workloads that do not require direct access to specialized hardware. In this paper, these nodes can be described as \emph{commodity-hardware VMs}, since they provide cloud-like elasticity on top of standard server infrastructure.

\paragraph{VMware VMs}

The second flavour consists of virtual machines provisioned on \textbf{VMware} infrastructure. Functionally, these nodes play a role similar to SUSE Virtualization VMs, but they are backed by a different virtualization stack. This flavour is useful where existing enterprise VMware infrastructure is already available or where operational constraints make VMware the preferred execution environment. It allows the platform to integrate with pre-existing virtualized estates while preserving the same Kubernetes management and GitOps workflow.

\paragraph{Commodity Bare-Metal Nodes}

The third flavour consists of \textbf{commodity bare-metal servers}. These nodes are used when workloads require direct access to local disks, NICs, GPUs, or other hardware components, or when virtualization overhead must be minimized. They are particularly appropriate for storage-intensive workloads, device-sensitive services, and performance-critical applications that benefit from predictable hardware behavior. Compared with VM-based workers, bare-metal nodes require a more explicit provisioning process, but they offer greater control over the runtime environment.

\paragraph{HPC Nodes from Alps}

The fourth flavour consists of \textbf{HPC nodes from Alps}, the research infrastructure operated by the Swiss National Supercomputing Centre (CSCS). Alps is a general-purpose compute and data infrastructure designed for extreme-scale scientific computing and AI. It supports the creation of \emph{vClusters} \cite{vclusters_24, vclusters_25}, which are logical partitions of the supercomputing resources dedicated to specific platforms or user communities.

A key advantage of Alps is its heterogeneous node portfolio. Depending on the allocation, Alps nodes may provide large multicore CPU configurations, GPU accelerators, or ARM-based architectures. This diversity makes Alps particularly attractive for workloads ranging from massively parallel CPU jobs to GPU-accelerated AI training and ARM-based experimentation.

\subsubsection{Kubernetes Cluster Assembly Models}

Using these node flavours, Kubernetes clusters are assembled in three main ways.

\begin{figure*}[h]
    \centering
    \includegraphics[width=\textwidth]{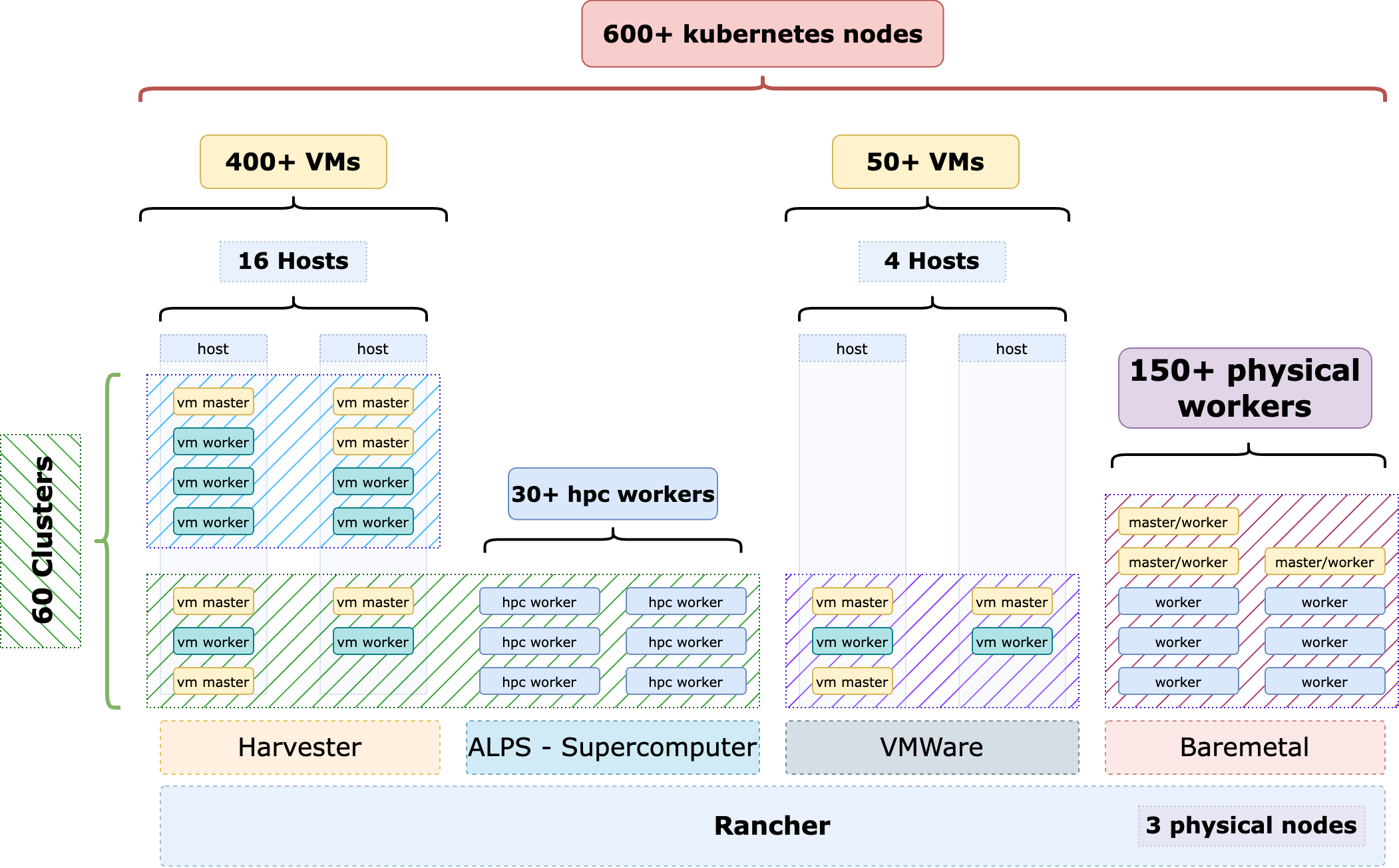}
    \caption{Kubernetes tenant distribution over host node types.}
    \label{fig:overview_all}
\end{figure*}

\paragraph{Fully Virtual Clusters}

The simplest model is a \textbf{fully virtual Kubernetes cluster}, where both control-plane and worker nodes run as VMs on SUSE Virtualization or VMware. This is the preferred model for standard platform services and moderately demanding application workloads. It provides fast provisioning, operational simplicity, and strong reproducibility. These clusters are typically used for web user interfaces, APIs, databases, middleware, development platforms, and other services that benefit more from elasticity than from direct hardware access.

\paragraph{Bare-Metal Clusters}

The second model is a \textbf{bare-metal Kubernetes cluster}. This model is used for heavier workloads that require direct access to disks, GPUs, or other hardware components, or where low-level tuning and maximum performance are important. Nodes in this category are provisioned through \textbf{MAAS} \cite{maas}, and in the future may also be provisioned through \textbf{OpenCHAMI} \cite{openchami_platform}.

Once the bare-metal nodes are installed and reachable, an \textbf{Ansible} playbook performs the operating-system configuration and prepares them to join the target Kubernetes cluster. In this architecture, OpenTofu remains responsible for generating the declarative cluster definition, while Ansible completes the host-side configuration and enrollment. This split keeps infrastructure definition centralized in Git while preserving the flexibility required for heterogeneous hardware provisioning.

\paragraph{Hybrid Alpernetes Clusters}

The third model is a hybrid architecture referred to as \textbf{Alpernetes}. In this design, a single Kubernetes cluster combines \textbf{SUSE Virtualization VMs} and \textbf{Alps HPC nodes}. The master nodes and a baseline set of worker nodes run on the virtualized platform, while additional worker nodes are provided by Alps to execute compute-intensive workloads. The result is a Kubernetes cluster that merges the operational stability of VM-based control infrastructure with the computational power of HPC hardware.

This model is especially useful for AI- and HPC-oriented use cases where user-facing services, workflow engines, gateways, or platform agents can remain on persistent virtual infrastructure, while high-demand execution is offloaded to specialized HPC nodes. It also allows the same cluster to expose conventional cloud-native services and high-performance execution backends without enforcing a strict separation between the two worlds.

\subsubsection{Integration of Alps Nodes into Kubernetes}

The Alps-backed worker nodes introduce specific operational constraints. These nodes are \textbf{diskless}, which means that each reboot effectively recreates the node from a clean state. For this reason, the integration workflow must assume that node-local state is ephemeral.

To use such resources, an \textbf{HPC cluster on Alps} must first be defined in the form of a \textbf{CSCS vCluster} \cite{vclusters_24, vclusters_25}. This vCluster is placed in a specific \textbf{VLAN on the high-speed network (HSN)}. Network-level connectivity must then be established so that the HPC nodes can communicate with the VM-based portion of the Kubernetes cluster. In practice, this requires opening the necessary \textbf{ACLs} to satisfy CNI communication and Kubernetes control-plane networking requirements.

After boot, an \textbf{Ansible} playbook configures each Alps node with the system prerequisites required to behave as a Kubernetes worker. This includes OS-level configuration, bind mounts to avoid problematic \texttt{overlayfs}-on-\texttt{overlayfs} situations, and filesystem preparation compatible with container runtimes and Kubernetes workloads. In our design, two main storage options are considered for these nodes: a \textbf{RAM-disk-based approach}, which provides the best performance but is fully ephemeral, and a \textbf{Ceph RBD-backed approach}, which offers more consistency at the cost of reduced performance.

Ansible is also responsible for joining the Alps node to the correct Kubernetes cluster. Because nodes are re-created on reboot, the automation must manage lifecycle corner cases explicitly. This includes removing and re-adding nodes in Rancher after a reboot, as well as detaching a node from one Kubernetes cluster and attaching it to another without rebooting. This capability is particularly useful for rapidly validating new platform releases on development or staging clusters before promoting them to production.

A key property of the design is that \textbf{any HPC node can be attached to any Kubernetes cluster}, provided that the corresponding network, provisioning, and authorization conditions are met. This makes the platform highly flexible: HPC capacity can be reassigned dynamically to different clusters according to testing needs, operational priorities, or changing workload demands.

\subsubsection{Summary}

Overall, the infrastructure combines \textbf{centralized Kubernetes governance}, \textbf{GitOps-based delivery}, \textbf{infrastructure-as-code}, and \textbf{heterogeneous compute backends} into a single private-cloud platform for HPC and AI. SUSE Rancher provides the management plane, SUSE Virtualization and VMware provide elastic VM-based resources, commodity bare-metal servers support performance-sensitive workloads, and Alps contributes specialized HPC capacity. OpenTofu and Terragrunt provide the declarative infrastructure layer, while ArgoCD ensures consistent platform deployment and reconciliation across all environments. The resulting architecture enables reproducible cluster provisioning while remaining flexible enough to incorporate both conventional cloud workloads and advanced HPC/AI execution resources.

\subsection{Self-Managed Sandboxes}
\label{sec:self_managed_sandboxes}
The inference service is designed to allow users to deploy arbitrary workloads in the form of namespace-isolated sandboxes. This approach supports multiple projects within the same cluster while providing flexible access patterns for different user needs.

Sandbox projects are defined in the CSCS Resource Management Portal (Waldur) \cite{waldur_platform}, which acts as the source of truth for project definitions, including project member management and resource quotas. OpenTofu continuously reconciles the desired state declared in Waldur by provisioning the necessary infrastructure components for each project: it creates the corresponding Git repositories, configures ArgoCD projects and their associated repositories, and deploys a top-level ArgoCD application per project to leverage the App of Apps pattern. This allows each project's applications to be managed hierarchically, with ArgoCD autonomously reconciling the desired state from the Git repository into the Kubernetes cluster. 

Identity Access Management informations — such as group membership — are used to configure authorization across the platform, ensuring that access to Git repositories, ArgoCD, and the Kubernetes API is correctly scoped to the members of each project.
Once a project is provisioned, its members can interact with the preconfigured Git repository to deploy their desired software stacks. By contributing to this repository, users can manage their workloads declaratively, with ArgoCD continuously syncing the repository state into their dedicated namespace. Workload status and health can be monitored through the ArgoCD web interface, while direct access to the Kubernetes API is available under strict RBAC policies that limit users to their own project resources.

Authentication across all components is unified through Dex, a federated OpenID Connect provider. Dex supports configuring a connector to an external Identity Provider (IdP), and OIDC clients can be created to manage access to other applications. In a first iteration, GitHub was configured as the IdP connector, enabling rapid development of the designed architecture. However, relying on an external IdP introduced challenges that proved difficult to address — for example, integrating with Waldur is problematic, as it relies on the CSCS IdP. In a second iteration, the plan is therefore to integrate CSCS's Keycloak-based IdP with Dex. However, moving away from GitHub comes with a trade-off: Git repository permissions can no longer be based on group membership in the CSCS IdP. To address this, an exposed instance hosting git repositories is expected to better integrate with the updated authentication and authorization design.

The underlying concept remains the same: a GitOps-ready platform that offers an intuitive and secure entry point to the inference service, allowing users to focus on developing, deploying, and running AI workflows.

\begin{figure*}[h]
    \centering
    \includegraphics[width=\textwidth]{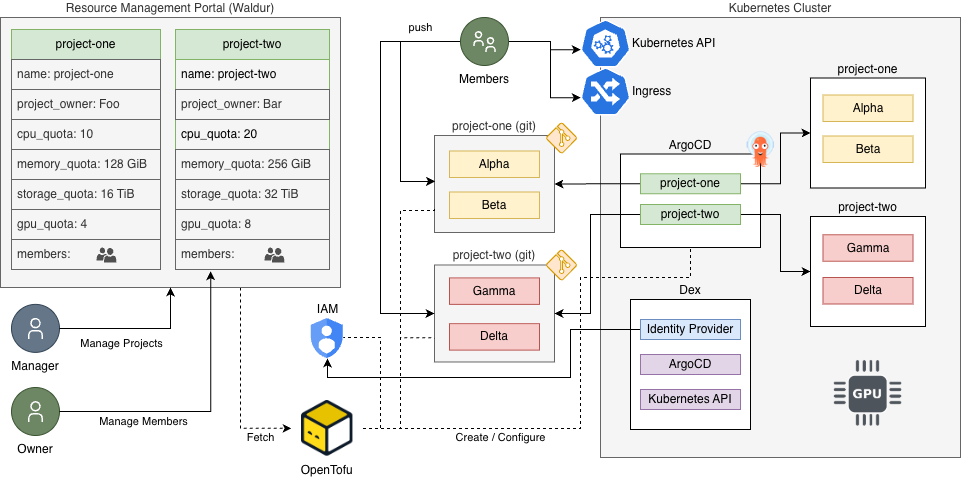}
    \caption{Sandbox provisioning and management leveraging IaC and GitOps methodologies.}
    \label{fig:sanboxes_provisioning}
\end{figure*}

\subsection{Fine-Tuning as a Service}
\label{sec:fine-tuning}

While pre-training establishes the latent capabilities of Foundation Models, the ``last mile'' of scientific utility is reached during a complex post-training adaptation lifecycle. This process is inherently non-linear and iterative. It begins with a data-centric phase where raw datasets are analyzed, curated, and filtered to create high-signal training sets. The workflow often initiates from a base model, potentially pre-trained within the same facility, and utilizes auxiliary models for synthetic data generation or automated labeling. The adaptation phase itself encompasses multiple sequential stages: Supervised Fine-Tuning (SFT) for instruction following, context-window extension for long-range dependencies, and Reinforcement Learning (RL) to align model outputs with project-centric preferences or domain-specific constraints. This is followed by release-critical optimizations, such as quantization or distillation into smaller architectures. Crucially, evaluation is not a terminal step but an activity interleaved between every stage to measure capability shifts.

\subsubsection{Catering to a Diverse User Base}
A primary driver for our service model is the emergence of the ``AI Antennas'' initiative, which aims to democratize HPC for a broader community, including SMEs and domain-specific research groups. We observe a clear dichotomy in our user base: expert HPC users who strongly-prefer low-level control via Slurm to tune bespoke architectures, and a new wave of users who wishes for ``one-click'' abstractions to integrate AI into their projects without managing the underlying ML design choices nor the infrastructure.

To bridge this gap, we focus on mitigating the risks of \textit{Catastrophic Forgetting} and \textit{Capability Collapse} \cite{understandingcatastrophicforgetting}. Fine-tuning can be a fragile process; misconfiguration can cause a model to lose the general reasoning, preferences, spatial awareness, acquired during pre-training. This risk is higher for larger models, where full-parameter updates can shift internal representations away from their optimized configuration. To protect the integrity of the base models, we are exploring curated, ``safe-by-default'' blueprint fine-tuning configurations (e.g., LoRA-based), shielding users from the underlying complexities. These can be composed in recipes catalogues for the most common cases.

\subsubsection{Bifurcated Orchestration: K8s and Slurm}
While some fine-tuning tasks can be executed on a few GPUs (for which a Kubernetes environment would be sufficient), scientific fine-tuning, particularly for high-resolution weather models, often demands scales that require the specialized interconnect performance. 

Deciding where a workload should reside (Kubernetes vs. Slurm) is not merely a matter of scale. Hosting fine-tuning logic concurrently on both workload managers would result in a costly duplication of engineering effort, for instance, in monitoring and debugging NCCL communications, and managing concurrent access to datasets, and similar. Consequently, we opted to concentrate the execution of fine-tuning workloads on Slurm while leveraging Kubernetes for the service-oriented frontend and inference.

We resolve this through a bifurcated orchestration strategy:
\begin{itemize}
    \item \textbf{The Control Plane (Kubernetes):} Manages the user-facing fine-tuning workflows, including dataset exploration and re-processing, evaluations, and experiment logging (e.g., via Kubeflow elements or MLflow).
    \item \textbf{The Execution Plane (Slurm):} Executes the actual compute-intensive training tasks. This allows us to re-utilize our mature and existing expertise in high-speed fabric optimization and GPU memory management immediately, avoiding the upfront duplication of effort required to stabilize these workloads in a cloud-native environment.
\end{itemize}

The bridge between these environments is \textit{FirecREST} \cite{palme2025firecrestv2lessonslearned}, a RESTful API layer developed at CSCS. FirecREST enables our K8s-centric services to programmatically submit and monitor jobs on the Slurm backend. Each FirecREST-submitted job relies on a predefined script, e.g. dedicated to a particular SFT activity. These scripts encapsulate the necessary environment variables, library paths, and communication tuning required for the \textit{Alps} fabric.

Under this model, users can compose complex model adaptation workflows by pulling these pre-defined building blocks from a curated catalog maintained by CSCS. This ``catalog-driven'' approach provides a middle ground for SMEs: they can customize high-level parameters (e.g., training epochs, rank of adaptation) while relying on pre-defined building blocks for the underlying hardware-specific optimizations.

While this bifurcated model represents our current design baseline, we are nonetheless building experience in running training and fine-tuning workloads directly on Kubernetes, specifically leveraging the Kubeflow suite. However, achieving the same level of confidence in performance and resource governance as our established Slurm environment will take time; the current strategy allows us to provide a high-quality service today while progressively developing the expertise required for a potential future switch to a fully cloud-native training environment.

\subsection{Managed Inference Service}
\label{sec:managed_inference_service}
This section describes the architecture and operational design of our inference service, which enables the controlled deployment and efficient operation of LLMs on HPC-oriented Kubernetes infrastructure. The system is built around declarative configuration, GitOps principles, and tight integration with existing HPC resource management and accounting workflows.

Model artifacts and deployment configurations are maintained in a repository accessible only to inference service administrators, providing a clear governance boundary between model consumers and platform operators. While the service has a broad catalog of models, additional models can be requested. New models onboarding requests undergo a vetting process. This evaluation assesses the model's computational footprint and projected operational costs, calculated on a per-token basis. This data is needed for configuring the \textit{Waldur} resource management portal \cite{waldur_platform} with appropriate quotas. Furthermore, the vetting phase informs capacity management requirements; specifically, for ``hot'' models requiring high availability, it ensures that sufficient compute capacity is reserved at secondary sites to support geo-redundant failover. Eventually, new models are onboarded through minimal YAML-based configuration, where parameters such as model source URLs, vLLM runtime settings, and tuning options are defined declaratively. This approach allows new inference backends to be deployed without modifying application code, improving reproducibility and operational consistency.

Deployment is managed using ArgoCD as a GitOps controller, which continuously reconciles the desired state stored in the repository with the runtime state of an RKE2 Kubernetes cluster. vLLM inference backends are deployed as Kubernetes workloads, with HPC nodes targeted through standard node labels and taints (e.g., \texttt{hpc=true}). This mechanism enables selective scheduling on specialized hardware while maintaining compatibility with general-purpose cluster resources.

At the access layer, LiteLLM \cite{litellm_2024} is used as a unified routing and control plane for LLMs inference. It provides centralized API access management, credential handling, and detailed usage metrics, including per-user and per-project consumption.
Once a model has deployed, it can be used to perform inference. Applications interact with the model through an API interface that accepts input data and returns the model’s generated output.
Access to the available models is managed through Waldur, which acts as the central platform for resource management, access control, and usage governance. Waldur \cite{waldur_platform} is an open-source platform commonly used to manage cloud and computational resources in research and enterprise environments. It provides capabilities such as user management, project organization, service provisioning, and tracking of resource consumption.
Within Waldur, administrators can generate and manage API keys that allow authorized users or applications to securely access the deployed models. These keys act as authentication credentials for making inference requests.
The inference gateway is implemented using LiteLLM, which is integrated with Waldur. LiteLLM acts as an abstraction layer between client applications and the underlying models, providing a unified interface for interacting with different model backends.
Through this integration, several governance and operational controls are available:
\begin{itemize}
    \item Authentication via API keys generated and managed in Waldur.
    \item Usage budgeting, allowing administrators to define limits on model usage or cost consumption.
    \item Rate limiting, which restricts the number of requests per user or per key to ensure fair and stable resource usage.
    \item Model selection, enabling access only to specific models depending on user permissions or project configuration.
\end{itemize}
The integration between Waldur and LiteLLM has been implemented using containerized services based on Docker. Each component runs within dedicated Docker containers, allowing the system to be deployed in a modular and reproducible way. This container-based architecture simplifies installation, scaling, and maintenance, while ensuring consistent environments across development, testing, and production deployments.
Overall, this setup provides a controlled and scalable mechanism for exposing trained models for inference while maintaining centralized management, security, and monitoring of resource usage.

\begin{figure*}[t]
    \centering
    \includegraphics[width=\textwidth]{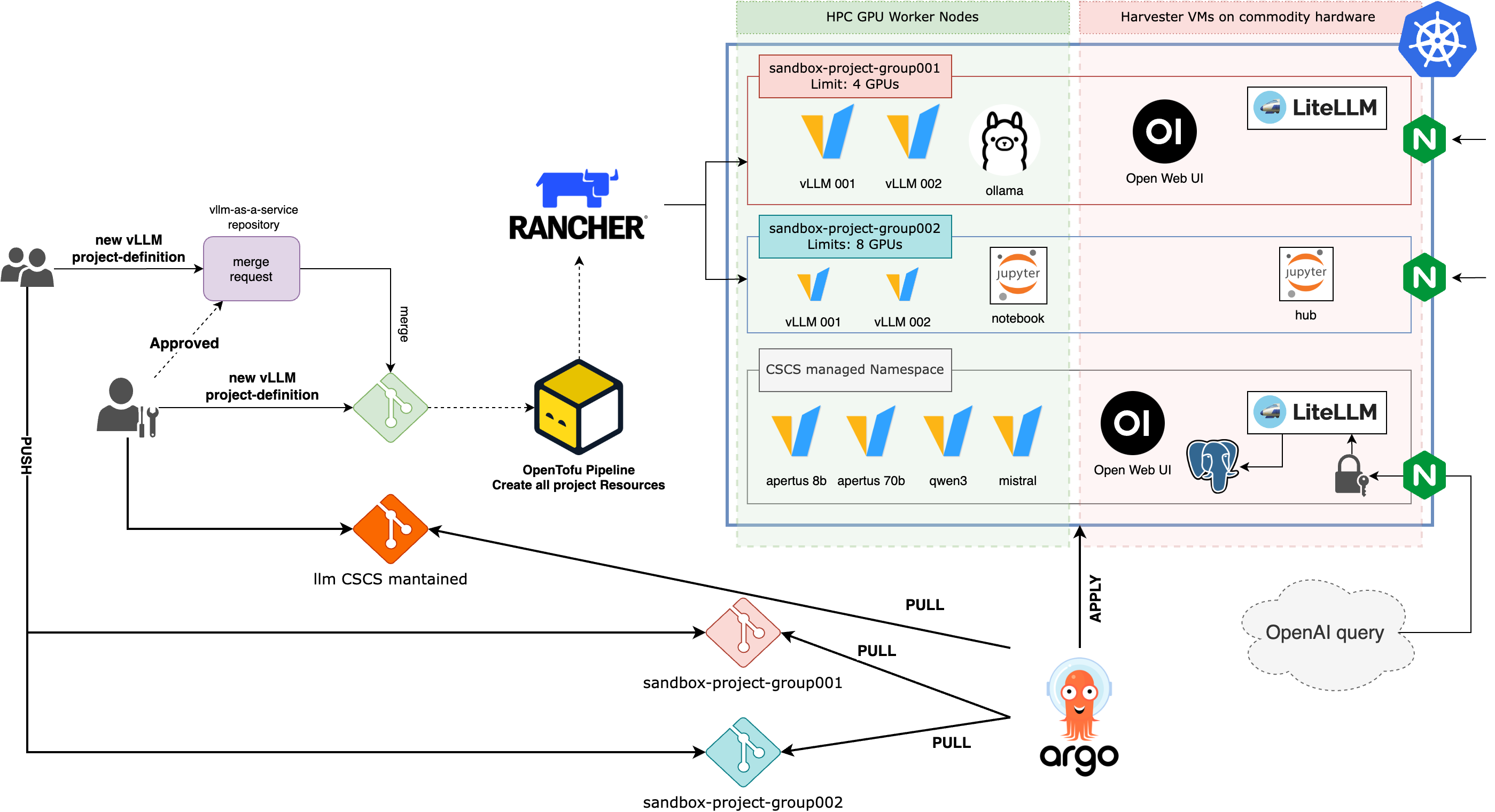}
    \caption{Alpernetes Cluster hosting inference service workloads}
    \label{fig:sys-arch}
\end{figure*}

\section{Results}
\label{sec:results}

The operationalization of the services described in this paper is an ongoing effort, with the current architecture representing an evolving solution design. While several technical aspects, such as the provisioning of common storage space between Slurm and Kubernetes, are not yet finalized, the system has reached sufficient maturity to support several concrete use-cases.

\subsection{Pilot Adoption and Use Cases}

The self-managed sandbox service has been adopted by pilot users, primarily members of the Swiss AI Initiative \cite{swiss_ai_initiative}. These researchers utilize the unified compute plane to deploy bespoke solutions (e.g., RAG stacks, coding assistants and agentic solutions) that require high-density GPU memory alongside traditional services (e.g., web interfaces, databses).

Simultaneously, the managed inference service has provided one of the backends for Public AI \cite{publicai_apertus_2025} since the launch of the Apertus model family in September 2025 \cite{apertus}. These workloads are focused on providing standardized API access to the Apertus-70B and Apertus-8B models \cite{apertus_huggingface_2025}, serving as one of our primary testbed for our managed inference service.

\subsection{Inference Performance Observations}

It is important to note that a formal, exhaustive performance analysis has not yet been performed. Our primary engineering efforts have been so far concentrated on the deployment and integration of the service layers, specifically operationalizing the quota and accounting flows between LiteLLM and Waldur. Consequently, we have not yet invested the necessary effort into the performance optimization of, e.g., the vLLM backends. 

The metrics observed over the past month are described below to offer readers a reference point for baseline performance on \textit{Alps} Grace-Hopper nodes, without the benefit of configuration tuning or specialized optimizations. We characterized the service by monitoring key LLM serving metrics: Queries Per Second (QPS), Time To First Token (TTFT), Inter-Token Latency (ITL), and End-to-End Latency (E2EL). Although we operate both 70B and 8B Apertus models, the following figures are not inclusive of speculative decoding related benefits.

\paragraph{Apertus-70B Metrics} 
The 70B parameter model exhibits a steady average ITL of $\sim$42ms. The TTFT remained consistently under 500ms for the majority of requests, with occasional P99 spikes reaching 2.5s. These spikes typically correlate with periods of high prompt length variability. During the monitoring period, prompt lengths for the 70B model were predominantly concentrated between 100 and 800 tokens, while response lengths remained moderate, typically within the 200 to 500 token range. The resulting average E2EL of 5.84s reflects the high computational intensity of a model of this scale when processing these generation lengths.

\paragraph{Apertus-8B Metrics} 
The smaller 8B model exhibits a significantly lower ITL average at $\sim$11ms. However, we observed a notably higher average E2EL of 31.4s during specific high-load intervals. An analysis of the request characteristics suggests a distinct usage pattern compared to the larger model: while prompt lengths remained short (often below 200 tokens), users frequently utilized the 8B model for long-form synthetic data generation, with individual responses often exceeding 3,000 tokens. These long-tail generation tasks push the limits of the allocated GPU memory and compute cycles, accounting for the increased end-to-end latency despite the superior per-token throughput.

\subsection{System Throughput and Reliability}

Token generation metrics provide a window into the daily operational load supported by the "Alpernetes" stack. Over a 48-hour period (March 15--16), the managed service facilitated the generation of approximately 2.5 million tokens for the 8B model and 1 million tokens for the 70B model.

\subsubsection{Uptime since September 5, 2025}
As described in the Hybrid Alpernetes Clusters section, this hybrid approach also provides significant reliability benefits. In HPC environments, supercomputers are inherently more prone to service interruptions caused by failures and scheduled maintenance. By hosting the control plane as well as lightweight services, such as databases, user interfaces, and other auxiliary components, on VMs outside the HPC system, the cluster can continue operating even when the HPC nodes are unavailable. In such cases, deployments remain active or pending until the HPC resources become available again, without losing their state. This preserves persistence and improves the overall resilience of the platform.

Since the cluster was created on September 5, 2025, it has not experienced any unplanned downtime, even though the HPC infrastructure has undergone several maintenance periods, including site-wide maintenance and upgrade operations.

\section{Ongoing and future work}
\label{sec:ongoing_future_work}

As part of ongoing and future work, the platform is being extended with cloud-native model management and serving capabilities. This includes the evaluation and integration of model registries for tracking model metadata and artifacts, as well as Kubernetes-native serving frameworks such as KServe. These components are currently being tested to enable consistent and reproducible deployment of both real-time and batch inference workloads using standard Kubernetes primitives and GitOps-based operational practices.

In parallel, we are refining the architectural design and operational model of the platform based on early adoption and testing feedback. Future iterations will consolidate these capabilities into a production-ready offering. The experience gained during this phase will inform best practices and provide practical guidance for supercomputing centers aiming to adopt Kubernetes-native AI workflows while preserving the operational constraints and performance expectations of traditional HPC environments.

\subsection{Unified Storage Access} \label{sec:unified_storage}

A critical aspect of enabling a seamless AI lifecycle across HPC and cloud-native environments is the availability of consistent and performant data access mechanisms from both Slurm-managed workloads and Kubernetes-orchestrated services. In practice, users must be able to access training datasets, intermediate artifacts, and model checkpoints transparently, regardless of whether workloads are executed in batch-oriented HPC jobs or long-running service deployments.

To address this requirement, we are currently investigating multiple storage integration strategies. One approach consists of mounting existing parallel file systems (e.g., scratch storage) directly onto Kubernetes nodes and exposing them to pods. This provides immediate compatibility with established HPC workflows and ensures high-throughput access to large datasets. However, this model introduces challenges in terms of access control, namespace isolation, and portability across clusters.

In parallel, the recent deployment of VAST storage systems introduces additional design opportunities. In particular, the availability of an S3-compatible object storage interface enables a more cloud-native access pattern, allowing data and model artifacts to be accessed uniformly from both Slurm and Kubernetes environments through standard APIs. This abstraction layer can decouple applications from underlying storage, improving portability and integration with existing tools.

At present, users often rely on external or semi-managed solutions for model and artifact storage, such as Hugging Face repositories or pre-existing CSCS-hosted object storage services. These approaches offer convenience and integration with common ML workflows, but they may introduce data locality and performance trade-offs when used at scale within HPC environments.

For sandboxed Kubernetes workloads, Ceph-backed Persistent Volume Claims (PVCs) are available and provide persistent, namespace-scoped storage. While suitable for many service-oriented workloads, these volumes are not optimized for the high-throughput access patterns typical of certain AI workloads.

Ongoing work aims to evaluate these options and converge towards a unified storage architecture that balances performance, usability, and reproducibility. A key objective is to provide users with a coherent data access model that spans both HPC and Kubernetes domains without imposing unnecessary complexity or compromising performance-critical workloads.

\subsection{Elastic Resources Management}
\label{sec:elasticity}

A fundamental challenge in operating inference services on HPC infrastructure is the mismatch between the dynamic, often bursty nature of service workloads and the static resource allocation model traditionally used in batch-oriented systems. Inference demand fluctuates over time, for example across diurnal cycles (e.g., lower load at night) or due to external events, causing underutilization when resources are statically provisioned for peak demand.

The current architecture relies on dedicating a subset of HPC nodes to a separate Kubernetes cluster for inference workloads. Although this approach provides clear operational boundaries and makes deploying services easier, it also causes a type of resource fragmentation that conflicts with the fundamental goals of HPC environments, where the focus is on maximizing overall utilization and minimizing idle resources. We acknowledge this as a current limitation of the platform.

To overcome this constraint, we are exploring approaches for introducing controlled elasticity into the system. In particular, we are evaluating the integration of OpenTela as a mechanism to dynamically manage a variable resource "delta" on top of a statically provisioned baseline.

In this model, a fixed pool of nodes remains permanently assigned to the inference-dedicated cluster. The size of this baseline is determined through empirical analysis, taking into account factors such as the number of models that must remain "hot" (i.e., pre-loaded in memory), expected concurrency levels, and service-level objectives.

On top of this baseline, an elastic layer of additional nodes can be dynamically allocated or reclaimed based on observed demand or predefined heuristics. This "delta" capacity enables the system to scale out during peak periods and contract during low utilization phases, thereby improving overall infrastructure efficiency while maintaining service responsiveness.

Several challenges remain in realizing this vision. These include defining robust scaling policies, ensuring fast and reliable node transitions between Slurm and Kubernetes environments, and minimizing the impact of such transitions on both batch workloads and long-running services. Furthermore, the interaction between elasticity mechanisms and data locality, particularly for large model weights, must be carefully considered to avoid performance degradation.

At this stage, this work is exploratory and serves to highlight an important open problem in the integration of HPC and service-oriented AI workloads. Achieving efficient elasticity without compromising the utilization and scheduling guarantees of HPC systems remains a key area of future investigation.

\subsection{High-Availability}

High Availability (HA) will be addressed in greater depth in future work. The primary objective is to ensure a defined level of service continuity during planned maintenance operations and in the presence of system faults.

In a conventional architecture, HA is typically achieved by duplicating infrastructure in a secondary data center. While this approach provides redundancy, it introduces a significant inefficiency: standby assets remain underutilized and are only activated during failure scenarios. This results in a cost model that is often economically unsustainable, particularly at scale.

To overcome these limitations, an active-active HA architecture is proposed. In an active-active configuration, multiple sites operate concurrently and serve production workloads simultaneously. This model significantly improves service continuity, as the system can react to failures in near real time without requiring manual intervention or failover procedures. Traffic can be dynamically redistributed across healthy nodes or sites, minimizing disruption and recovery time objectives (RTO).

However, the adoption of an active-active HA model introduces several technical challenges that must be carefully addressed. Among the most critical are:

\textbf{Split-brain scenarios}, where network partitions cause nodes or sites to operate independently, potentially leading to data inconsistency or conflicting states.

\textbf{Latency considerations}, particularly in geo-distributed environments, where inter-site communication delays can impact synchronization, consistency models, and overall application performance.

\textbf{State management and data replication}, which require robust strategies to ensure consistency across distributed systems.

One potential approach to implementing such an architecture is through a Kubernetes-based cluster mesh. In this model, multiple Kubernetes clusters—potentially distributed across different data centers or regions—are interconnected to form a unified control and service plane.

A cluster mesh enables:

\begin{itemize}
    \item Cross-cluster service discovery, allowing services running in different clusters to communicate seamlessly.
    \item Global load balancing, where traffic is intelligently routed to the closest or healthiest service endpoint.
    \item Workload distribution, enabling applications to run concurrently across multiple clusters.
    \item Resilience and fault isolation, as the failure of a single cluster does not compromise the entire system.
\end{itemize}

Technologies such as service meshes (e.g., Istio or Cilium) can extend this concept by providing advanced traffic management, observability, and security features across clusters. In particular, a mesh-enabled Kubernetes deployment can mitigate some of the challenges of active-active HA by handling failover, retry logic, and circuit breaking at the network layer.

Despite its advantages, a cluster mesh architecture also introduces additional complexity in terms of networking, security policies, and operational management. Careful design is required to balance consistency, availability, and performance, especially in high-performance computing (HPC) environments where latency and throughput are critical factors.

\subsection{DDP over HPE Slingshot on Kubernetes}
\label{sec:hsn-training}

To begin evaluating the performance of the Slingshot-11 interconnect within our Kubernetes-orchestrated environment, we adopted the CIFAR-10 image classifier toy use-case \cite{pytorch_cifar10_tutorial}, adapting it for Distributed Data Parallel (DDP). 

In this section, we present initial results from training this model across GH200 nodes over the Slingshot High-Speed Network (HSN) from Kubernetes-managed containers.

\subsubsection{Environment}

Experiments ran on the \textit{Forno} cluster (CSCS Alps), where each node
is a 4-way GH200 Superchip: 4$\times$ H100 GPUs (96\,GB HBM3), 288 ARM
cores, $\sim$487\,GB LPDDR5X, and 4$\times$ Slingshot-11 CXI NICs
(\texttt{hsn0}--\texttt{hsn3}, 200\,Gbps each). The stack uses RKE2~v1.33,
Cilium, NVIDIA GPU Operator, Kueue, and Kubeflow Training Operator~v2. No
Multus or RDMA device plugin is deployed; pods use \texttt{hostNetwork:~true}.

\subsubsection{Challenges}

Unlike Slurm, where \texttt{srun} natively provides the HSN fabric to all
ranks, Kubernetes required solving several issues:

\begin{itemize}
  \item \textbf{Master discovery:} With \texttt{hostNetwork}, pod IPs fall in
    the HSN subnet and collide with other host-network services (Ceph-CSI,
    Hubble), breaking headless-service DNS. We bypass DNS by querying the K8s
    API for rank-0's \texttt{nodeName} and resolving it via host DNS to obtain
    the \texttt{hsn0} IP.
  \item \textbf{Interface selection:} The K8s \texttt{InternalIP} maps to
    inconsistent HSN interfaces across nodes. We pin NCCL and the master
    address to \texttt{hsn0} explicitly.
  \item \textbf{CUDA compatibility:} The NGC container version must match the
    host CUDA driver; we use \texttt{pytorch:24.04-py3} for driver~12040.
  \item \textbf{Rendezvous semantics:} PyTorch propagates \texttt{MASTER\_ADDR}
    literally to all workers; using \texttt{0.0.0.0} caused cross-node timeouts.
  \item \textbf{NCCL transport:} No IB devices exist on Slingshot nodes. With
    \texttt{NCCL\_IB\_DISABLE=1}, NCCL uses TCP sockets over HSN. Native CXI
    RDMA requires the \texttt{aws-ofi-nccl} plugin compiled against Cray
    \texttt{libfabric}~1.15.2 (available on the host); this is in progress.
\end{itemize}

\subsubsection{Setup and Results}

We benchmark with ResNet-18 on CIFAR-10 (2~nodes, 4~GPUs each, world
size~8, batch~256/GPU, 50~epochs, NCCL~2.21.5). This small model isolates
network overhead from compute. Table~\ref{tab:hsn-perf} summarizes results.

\begin{table}[htbp]
\centering
\small
\caption{DDP training: 2 nodes $\times$ 4 GPUs, 50 epochs.
Initial results. Analysis and optimizations pending. Slurm-based runs included for comparative reference.}
\label{tab:hsn-perf}
\setlength{\tabcolsep}{3pt}
\begin{tabular}{@{}llrrr@{}}
\hline
\textbf{Env.} & \textbf{Network} & \textbf{Runtime\,(s)} & \textbf{Mean ms/iter} & \textbf{Speedup} \\
\hline
K8s     & eth0 (TCP)     & 3779 & 190  & 1.0$\times$  \\
K8s     & hsn0 (TCP)     & 1165 &  58  & 3.2$\times$  \\
K8s$^\dagger$ & hsn0--3 (TCP) & 1550 &  79  & 2.4$\times$  \\
K8s$^\dagger$ & CXI RDMA (OFI) & \multicolumn{3}{c}{\textit{work in progress}} \\
\hline
Slurm   & CXI RDMA       & 81     & 4     & ---          \\
\hline
\end{tabular}
\end{table}

Switching from the management network (overlay) (\texttt{eth0}) to a single HSN
interface yielded a \textbf{3.2$\times$} speedup. Using all four HSN
interfaces did not help: for ResNet-18's small allreduce payload
($\sim$45\,MB), the multi-socket overhead outweighed any bandwidth gain.
Intra-node communication used P2P/CUMEM over NVLink-C2C as expected.

\subsubsection{Ongoing Work}

These results are preliminary and not yet at parity with native Slurm.
Key remaining steps: (1)~building the \texttt{aws-ofi-nccl} plugin for CXI
RDMA with GPU Direct, (2)~populating the Slurm baseline rows,
(3)~testing at larger scale and with heavier models, and (4)~deploying
Ansible-prepared sysctl tuning and \texttt{hsn0} IP pinning to the nodes.

\subsection{vLLM and KubeRay Multi-Node Serving}

The vLLM \textit{production-stack} \cite{vllm_production_stack_2025} is already used for single-node model serving. This section describes the ongoing effort to extend it to \textbf{multi-node} inference using \textbf{KubeRay} for pipeline-parallel serving on GH200 nodes. Several earlier approaches were explored before converging on the current Helm-based setup: a raw \texttt{RayCluster} manifest, a \texttt{RayService}-based deployment, and an NVIDIA NIM LLM multi-node configuration (all preserved and still under evaluation). The NIM approach was discarded because the NIM container images for GLM-5 are not available for ARM (aarch64), which is required by the GH200 Grace CPU. This work is ongoing.

\subsubsection{Current Setup}

\begin{itemize}
  \item \textbf{Model:} GLM-5-FP8 (\texttt{zai-org/GLM-5-FP8})
  \item \textbf{Worker nodes:} 2$\times$ GH200 nodes, each with 4$\times$ NVIDIA GH200 120\,GB GPUs
  \item \textbf{Parallelism:} Pipeline Parallel (PP) = 2 $\rightarrow$ 1 head pod + 1 worker pod across 2 nodes; Tensor Parallel (TP) = 4 $\rightarrow$ all 4 GPUs within each node; Total GPUs = 8
  \item \textbf{Chart version:} vllm-stack \texttt{0.1.10}
\end{itemize}

\subsubsection{Known Issues \& Next Steps}

\begin{itemize}
  \item \textbf{Inter-node GPU communication is too slow over standard TCP/eth0.} NCCL currently falls back to socket transport (\texttt{NCCL\_IB\_DISABLE=1}), which causes inference timeouts and ``thinking failures'' during pipeline-parallel execution between the head and worker pods.
  \item \textbf{Next step: enable the libfabric CNI plugin} to expose high-performance fabric interfaces (e.g.\ CXI\,/\,Slingshot) directly into pods. This will allow NCCL to use native fabric transport instead of TCP sockets, drastically reducing inter-node latency and eliminating the communication-related failures.
  \item Additional tuning still required: adjust \texttt{gpuMemoryUtilization}, \texttt{maxModelLen}, and \texttt{maxNumSeqs} based on production load; re-evaluate DeepGEMM (\texttt{VLLM\_USE\_DEEP\_GEMM}) once the OOM during FP8 requantization is resolved upstream; consider enabling Ray compiled DAG (\texttt{VLLM\_USE\_RAY\_COMPILED\_DAG=1}) after the network path is upgraded.
\end{itemize}

\subsection{Services for Data and Artifact Governance}
\label{sec:data_governance_services}

The iterative nature of fine-tuning activities, driven by potentially large research teams exploring combinations of techniques, parameters, and data mixtures, introduces a ``checkpoint explosion'' risk. While a single 70B parameter model checkpoint consumes $\sim$140\,GB, the iterative process can generate a large number of these artifacts across development cycles. A similar consideration can be made for datasets, as researchers produce multiple versions while experimenting with different mixtures and filtering strategies. Together, these can lead to petabyte-scale storage exhaustion and complicates the lineage tracking needed for reproducibility and legal compliance.

Traditional file-system structures and ACL-based policies can be insufficient tools. Reclaiming storage space (a need that can arise suddenly for many reasons, given the typically shared nature of HPC storage systems) can be a precarious activity given that during the course of a project it can become unclear which data (e.g., checkpoints or re-processed datasets) are needed for future reproducibility and which can be safely deleted.

We recognize the need for an integrated MLOps metadata layer to automate artifact lifecycle management and provenance tracking. To this end, we plan to explore the integration of open-source tools into our ML platform. This prospective governance framework would be offered to PIs as an optional service, providing structured artifact management while retaining as much as possible of the current flexibility.

\subsection{Trusted Computing Environment}
\label{sec:inference_in_trusted_research_environments}

A key frontier for expanding the platform's impact is its application to sensitive data use-cases, most notably in medical and clinical research where datasets may contain non-anonymized patient records subject to strict regulatory frameworks. Unlike the general-purpose sandbox model described in \hyperref[sec:self_managed_sandboxes]{Section~\ref*{sec:self_managed_sandboxes}}, these workloads impose a qualitatively different set of requirements: the infrastructure provider (CSCS) must not have access to the data at rest or in transit, placing the data sovereignty boundary firmly at the user organization level.

To satisfy these constraints while maintaining a clear legal and operational separation of responsibilities, Trusted Computing Environments (TCEs) are realized strictly on top of an Infrastructure-as-a-Service (IaaS) model. In this paradigm, CSCS provides isolated compute, storage, and networking primitives, but does not operate or manage any platform-level services within the TCE. Instead, each participating institution is responsible for deploying, configuring, and operating its own complete software stack, including identity and access management, orchestration, and inference services. This approach ensures that CSCS acts solely as a data processor of encrypted infrastructure resources, without visibility into or control over the sensitive workloads executed within the environment.

From an architectural perspective, a TCE is instantiated as a dedicated Alpernetes cluster, isolated at the network layer through a dedicated VLAN with strict ACL enforcement preventing lateral connectivity to any other tenant or platform service. Storage is provisioned through encrypted volumes where encryption keys are exclusively held and managed by the data-owning institution, ensuring that neither platform operators nor other tenants can access the underlying data. All traffic ingressing and egressing the TCE perimeter is encrypted end-to-end, preserving confidentiality even within the shared physical fabric.

Within this boundary, institutions may deploy inference services following the same architectural principles as the managed inference stack described in \hyperref[sec:managed_inference_service]{Section~\ref*{sec:managed_inference_service}}, including the use of HPC-grade GPU nodes and standardized API interfaces. However, unlike the managed offering, all components of the stack, including model serving backends, orchestration layers, and access control mechanisms, are instantiated and operated by the institution itself. This is particularly critical as models may be fine-tuned on sensitive domain data, making the ability to host private, institution-specific models a first-class requirement.

A key design implication of this approach is that all platform components described in this paper, such as sandbox environments, fine-tuning pipelines, and inference services, must be designed as fully reproducible, portable, and externally deployable building blocks. Through GitOps and Infrastructure-as-Code (IaC) practices, we ensure that the same reference architectures used internally can be shared with external institutions, enabling them to redeploy identical or customized stacks within their own TCEs. This establishes a consistent operational model across managed and self-managed environments, while preserving institutional autonomy and compliance with strict data governance requirements.

This work represents a concrete instantiation of the privacy-preserving, air-gapped environments introduced in \hyperref[sec:agility_through_autonomy]{Section~\ref*{sec:agility_through_autonomy}}. It enables institutions to perform inference over sensitive data without exporting it outside the CSCS infrastructure, while maintaining full control over both data and software lifecycle within a legally and operationally well-defined boundary.

An external assessment and formal certification process of this approach are currently pending, and will be required to validate its compliance with applicable regulatory and security standards before it can be considered operational for production-sensitive workloads.

\section{Discussion}
\label{sec:discussion}

Ultimately, providing end-to-end support for the AI lifecycle is becoming a strategic necessity for national facilities to enable the development of sovereign AI capabilities. Our investigation demonstrates that the friction between batch-oriented HPC and service-oriented AI is primarily an engineering challenge. The presented hybrid architecture supports diverse stateful services, ranging from lightweight RESTful APIs to GPU-intensive applications, while providing the governance framework necessary for users to remain agile in a highly competitive and rapidly evolving field. Realistic inference scenarios and empirical data are needed for informed decisions on compute planning (w.r.t., pre-fill and decode phases), although we note that a systematic performance characterization and fine-tuned configuration of inference backends remains part of future work.

The observed performance gap between Kubernetes-based workloads and native Slurm executions, despite leveraging the Slingshot high-speed network, can be largely attributed to the maturity of the Slurm execution stack, which benefits from extensive, production-grade performance engineering, whereas the Kubernetes-based path is for us still in an early stage where comparable optimization effort has not yet been invested. From an engineering perspective, this directly motivates a bifurcated orchestration model, where Kubernetes acts as a flexible service control plane while Slurm remains the execution backend for tightly coupled, performance-critical workloads.

From an operational standpoint, these early results suggest that the traditional distinction between HPC and cloud models could be revisited as an extension of the current paradigm, rather than as a replacement for it. HPC systems have historically been optimized for capability workloads, characterized by tightly coupled, large-scale computations, while cloud platforms are designed for service workloads, emphasizing elasticity and long-running applications. The AI lifecycle inherently spans both paradigms, and inference workloads, when operating at scale with high concurrency, large context windows, and strong data locality requirements, can themselves exhibit capability-like characteristics. This reframing highlights that hybrid architectures are not transitional solutions, but a necessary evolution to support emerging scientific and industrial AI workloads.

We also emphasize that the conclusions drawn in this work are bounded by the current system maturity and evaluation scope. Our results are based on early-stage deployments and limited-scale experiments, and do not yet generalize to fully optimized systems or alternative hardware and network configurations. In particular, different interconnect technologies, storage systems, or workload characteristics may lead to different trade-offs, and smaller-scale or loosely coupled inference workloads may still be more efficiently served on conventional cloud infrastructure. Nevertheless, to the best of our knowledge, we did not identify directly comparable hybrid end-to-end architectures in the existing literature; we therefore believe that, despite its current limitations, this initial investigative work provides a concrete and practical blueprint for other institutions seeking to operationalize the full AI lifecycle on HPC infrastructure.

\section{Conclusions}
\label{sec:conclusions}

This paper presented our ongoing effort to operationalize the \emph{full lifecycle} of Foundation Models within a national supercomputing facility. Starting from the observation that pre-training alone is insufficient to realize the goals of sovereign AI, we described how CSCS is extending its traditional HPC service model toward a hybrid architecture capable of supporting the setup of custom services, execute post-training adaptation and long-running inference workflows in addition to large-scale batch computation. Concretely, we introduced \textit{Alpernetes}, a Kubernetes-managed platform that combines virtualized commodity hardware (VMs) with \textit{Alps} HPE Cray EX nodes into a unified compute environment for service-oriented AI workloads.

On top of this architecture, we outlined three complementary service models. First, \emph{self-managed sandboxes} provide namespace-isolated environments in which users can autonomously deploy custom AI services and broader software stacks, including non-LLM workloads, while benefiting from GitOps-based operations and a heterogeneous compute plane. Second, we described our current \emph{fine-tuning approach}, which adopts a bifurcated orchestration model: Kubernetes hosts the user-facing control plane and higher-level workflow logic (primarily for non-expert users), while Slurm remains the execution backend for the underlying compute-intensive tasks. This design allows us to preserve the maturity and performance engineering of the HPC stack while progressively building more accessible abstractions for a broader user community. Third, we presented our \emph{managed inference service}, where declarative onboarding, gateway-based access control, Waldur-based governance, and vLLM backends on HPC nodes together provide a controlled and scalable mechanism for exposing inference APIs.

Although the platform is still under active development, the initial results already show that this direction is both technically feasible and operationally meaningful. Pilot users are actively using the platform to build custom AI workflows, managed inference has supported production-facing workloads (primarily coming from externally hoster web UIs), and our early experiments demonstrate that Kubernetes-orchestrated workloads can exploit the Slingshot high-speed network, albeit not yet at the level of native Slurm. These findings clarify both the promise and the current limitations of the approach: hybrid orchestration expands the usefulness of national HPC facilities for modern AI workflows, but achieving performance parity for tightly coupled workloads still requires substantial engineering effort.

After closing this paper, the reader should remember one central point: supporting sovereign AI at national facilities cannot stop at pre-training. The real value of Foundation Models is realized across a broader lifecycle that includes adaptation, serving, governance, and operational continuity, and these phases do not fit cleanly within the historical boundaries of either traditional HPC or conventional cloud alone. Our work suggests that the right response is not to replace one model with the other, but to combine them deliberately: preserving HPC where capability-oriented execution is essential, while introducing cloud-native service layers where agility, autonomy, and persistent operation are required. Even at this early stage, we believe this hybrid design provides a concrete blueprint for how supercomputing centres can evolve to support the next generation of scientific and industrial AI workloads.

\begin{acks}
We thank our user community for their valuable contributions, especially members of the Swiss AI Initiative, whose collaboration and feedback are crucial in shaping our solutions.

We acknowledge using tools such as Writefull, ChatGPT, and Gemini to improve the clarity and readability of this document. All suggestions from these tools were carefully reviewed and/or revised by the authors to preserve the original intent. The tools were used solely for language editing, not for generating ideas or data.
\end{acks}

\bibliographystyle{ACM-Reference-Format}
\bibliography{resources}

\end{document}